%% file: ProcAmes.tex
\documentclass[a4paper,10pt]{article}

\input{headers}
\usepackage{a4wide}
\usepackage[usenames]{color}
\input{commands}

\newcommand{\clr}[1]
{
{\color{Black}{#1}}
}

\title{
\clr{An extinction rule for a class of 1D quasicrystals}
}
\author{
\textsc{Pawe\l{} Buczek} 
\footnote{Current address: Max Planck Institute of Microstructure Physics, Weinberg 2, D-06120 Halle, Germany (also address for correspondence).}
\footnote{Electronic address: \texttt{\small{pbuczek@mpi-halle.mpg.de}}} \textsc{and}
\textsc{Janusz Wolny}
\footnote{Electronic address: \texttt{\small{wolny@novell.ftj.agh.edu.pl}}} \\
\textit{\small{Faculty of Physics and Applied Computer Science, AGH-UST}} \\ 
\textit{\small{al. Mickiewicza 30, 30-059 Krak\'ow, Poland}} 
}
\date{\today}

\begin{document}
\maketitle

\begin{abstract}
\clr{We study decorated one-dimensional quasicrystal obtained by a non-standard projection of a part of two-dimensional lattice. We focus on the impact of varying relative positions of decorated sites. First, we give general expression for the structure factor. Subsequently we analyze an example of extinction rule.}

\end{abstract}

\section{Introduction}
\label{sec:Introduction}

\clr{Two decades after discovery of quasicrystals studies of their decorations are still a challenge for crystallographers. The ``cut-and-project'' scheme can provide us with a structure factor for an (almost) arbitrarily decorated model set. Nevertheless, this technique often appears to be simply inconvenient, especially when it requires utilization of fractal atomic surfaces.

There exists another method, which allows for simplification of calculations of quasicrystalline spectra, especially in case of decorated structures. The method is based on the concept of reference lattice, \cite{Wolny1998PhM}, which focuses on the statistical description of spatial properties of atomic orderings. In this method each type of decorating atoms manifests itself as an additional component of the so-called displacement density function, which constitutes a so-called average unit cell, \cite{Buczek2005b}.

This paper may be regarded as an extension of \cite{Buczek2005b}. We discuss the impact of varying decoration on Fourier spectrum of one-dimensional quasicrystal. First, the average unit cell technique is briefly discussed. Subsequently, we describe the model of the structure (Section \ref{sec:TheModel}) and study the decorations (Section \ref{sec:TheImpactOfDecorations}). We explicitly calculate the structure factor and analyze an example of extinction rule.
}

Let us summarize briefly the average unit cell technique. Let $\Lambda \subset \Reals^{n}$ be Delone set and $\Lambda'=\left\{ x_{l} \right\}$ its orthogonal projection on the scattering vector $\vec{k}=\frac{2\pi}{\lambda} \widehat{\vec{k}}$. Let us assume that all the atoms have identical scattering powers $f$; we will release this assumption soon. Displacement sequence $u_{l} = x_{l} - \left\| x_{l}/\lambda \right\|\lambda \in \left[-\lambda/2,\lambda/2 \right]$, where $\left\| x \right\| = \floor(x + \frac{1}{2})$ is the nearest integer function, usually has a well defined statistical distribution $P_{\lambda}(u)$. We call $u$ the displacement variable associated with the scattering vector $\vec{k}$. The $\vec{k}$-mode of its Fourier transform, $F(\vec{k}) = f \int_{-\infty}^{\infty} P_{\lambda}(u) \exp(\ii k u) d u$, gives the value of structure factor for scattering vector $\vec{k} \in \Reals^{n}$.

In general a separate distribution should be created for each $\vec{k}$, but in the case when diffraction pattern is supported on the set of the form
\begin{align}
\mathcal{R} = \left\{\sum_{s=1}^{m} h_{s} \vec{K}_{s}|\,h_{s} \in \Zahl,\,\vec{K}_{s} \in \Reals^{n} \right\}
\end{align}
the task may be \clr{considerably} simplified. We need to compute the joint probability $P(\vec{u})$, where $s$-component of $\vec{u}$, $\left(\vec{u}\right)_{s}$, stands for the displacement variable pertaining to reference lattice induced by vector $\vec{K}_{s}$. The structure factor can be then computed as
\begin{align}
	F(\vec{k}) = f \int_{R^{m}} P(\vec{u})\exp(\ii\vec{k}\cdot\vec{u}) d^{m}\vec{u},
\end{align}
providing $\vec{k}\in\mathcal{R}$.

The task task gets slightly comlicated, when we are to deal with decorated structures. We introduce the notion of decoration sites, \cite{Buczek2005b}. Decoration sites form a skeleton of the atomic structure. To each decoration site we assign a group of decorating atoms. The decoration sites are divided into subsets; we say that each decorating site in a given subset is of certain common type (the types are denoted by the first capital letters of alphabet: $A$, $B$, $C\ldots$). It is a generalization of classical terms -- in traditional crystallography each crystal has only one type of decoration sites, organized in one of the Bravais lattice. The number and types of the atoms assigned to the site depend only of its type. The number of atoms assigned to site $X$ is denoted by $n_{X}$, $X = A,\,B,\,C\ldots$, and $\sum_{X} n_{X} = N$, $N$ being the total number of different decorating atoms. The relative number of the given type of sites in the structure is denoted by $\varphi_{X}$; these numbers are normalized: $\sum_{X} \varphi_{X} = 1$. $f_{X_{l}}$ represents the scattering power of $l^{\mathrm{th}}$ atom in the group of decorating atoms corresponding to a site of type $X$. The position of the atom with respect to the site is $\Delta_{X_{j}} \in \Reals^{n}$. We will use the notation $a^{s}_{X_{l}} = \Delta_{X_{l}}\cdot\widehat{\vec{K}}_{s}$. When there is no danger of confusion (e.g.~for one-dimensional structures) we will skip the index $s$. Overall displacement density function for our structure reads
\begin{align}
	P(\vec{u}) = \frac{1}{\mathfrak{N}}
		\sum_{X}\varphi_{X}\sum_{l=1}^{n_{X}} f_{X_{l}} \mathcal{P}_{X}(\vec{u} - \vec{a}_{X_{l}}),
\end{align}
where $\mathcal{P}_{X}$ is the joint displacement probability for the reference sites of type $X$ and the normalization reads	$\mathfrak{N} = \sum_{X} \sum_{l=1}^{n_{X}} f_{X_{l}} \varphi_{X}$. $\vec{a}_{X_{l}}\in\Reals^{n}$ stands for the vector which components are $a^{s}_{X_{l}}$. The structure factor for $\vec{k}\in\mathcal{R}$ is
\begin{align}
	F(\vec{k}) = \mathfrak{N} \int_{R^{m}} P(\vec{u})\exp(\ii\vec{k}\cdot\vec{u}) d^{m}\vec{u}.
\end{align}
Please refer to \cite{Buczek2005b} for more detailed exposition and discussion.

\section{The model of structure}
\label{sec:TheModel}

To visualize the impact of the above processes we will work with a simple one-dimensional quasicrystalline ordering; there are no obstacles to repeat the approach for the higher dimensional structures \clr{(see e.g.~\cite{Kozakowski2004}, where the approach was successfully used in 2D)}. Our structure is built upon Sturmian sequence of two spacings of lengths $\mathcal{A}$ and $\mathcal{B}$. The reference sites are placed on the joints between the segments. Site is of type $X$ (either $A$ or $B$) if the next site is in the distance $\mathcal{X}$ ($\mathcal{X}=\mathcal{A},\,\mathcal{B}$). Set of decoration sites
\begin{align}
\Lambda_{0} = \left\{x_{m} = \left\|m/(\sigma+1)\right\|\mathcal{A} + (m - \left\| m/(\sigma+1) \right\|) \mathcal{B} \right\},
\end{align} 
where $\sigma\in\Reals\setminus\mathbb{Q}$. \clr{The set can be obtained using the ``cut-and-project'' method in a two-stage non-standard scheme. First, part of a square lattice is projected onto a line with slope $\sigma$ \cite{Senechal1995,deBruijn1981}. Subsequently, the two interatomic lengths $\mathcal{A}$ and $\mathcal{B}$ present in the structure are adjusted (the adjusting process is meant to keep the atomic density constant).} We set \clr{$\mathcal{A} = \frac{\kappa(1+\sigma)}{\kappa+\sigma}a_{0}$, $a_{0} = 1 + 1/\tau^{2}$, $\tau$ being golden ratio, and $\mathcal{B} = \mathcal{A}/\kappa$}, $\kappa\in \Reals_{+}\setminus\left\{1\right\}$. $\sigma$ determines the sequence of sites and $\kappa$ the relative length of spacings. Easy calculations show that the relative concentrations of sites are $\varphi_{A} = 1/(1+\sigma)$, $\varphi_{B} = 1 - \varphi_{A}$.

$\Lambda_{0}$ is incommensurately modulated crystal; its spectrum contains two incommensurate vectors, \cite{Yamamoto1982}, $k_{0} = 2\pi/a_{0}$ and $q_{0} = 2\pi/b_{0}$, $b_{0} = (\sigma + 1)a_{0}$. The two dimensional displacement density function for the atom $f_{X_{l}}$ induced by these two vectors reads
\begin{align}
	P_{X_{l}}(u,v) = \frac{1}{U_{X}} 
		\delta(v - \xi u - a_{X_{l}}(1-\xi))R(u;U_{X},c_{X} + a_{X_{l}}),
\end{align}
where $\xi = (1+\sigma)/(1-\kappa)$, $U_{X} = (\mathcal{A} - \mathcal{B}) \varphi_{X}$, $c_{A} = - (\mathcal{A} - \mathcal{B}) \sigma/(2(1+\sigma))$ and $c_{B} = - c_{A}/\sigma$. $R(x;w,x_{0})$ is the rectangle function with center $x_{0}$, width $w$ and height 1. $\delta(x)$ stands for Dirac's delta function. The overall density function is the weighted sum of such functions, $P(u,v) = \mathfrak{N}^{-1} \sum_{X}\sum_{l=1}^{n_{X}} f_{X_{l}} \varphi_{X} P_{X_{l}}(u,v)$. The structure factor for $k=n_{1} k_{0} + n_{2}q_{0}$ is
\begin{align}
	F(k) = \sum_{X}\sum_{l=1}^{n_{X}} f_{X_{l}} \varphi_{X} F_{X_{l}},
	\label{eq:GeneralStructureFactor}
\end{align}
where
\begin{align}
	F_{X_{l}} &= \exp(\ii(K c_{X} + k a_{X_{l}}))\sin\fbr{K U_{X}/2}/\fbr{K U_{X}/2},
		\label{eq:PartialFactor} \\
	K &= k - n_{2}q_{s}, \label{eq:K} \\
	q_{s} &= q_{0}(1-\xi) \label{eq:qs}.
\end{align}
$F_{X_{l}}$ is partial structure factor associated with given type of decorating atoms. $q_{s}$ will be referred to as {\it shift vector}. It is the basic quantity for determining of the extinction rules presented in Section \ref{sec:TheImpactOfDecorations}.

\clr{Fibonacci crystal \cite{Senechal1995} can be obtained in the above formalism for $\sigma = 1/\tau$ and $\kappa = \tau$, where $\tau$ stands for golden ratio.}

%
%
%

\section{Impact of decorations}
\label{sec:TheImpactOfDecorations}


\clr{Let us study the impact of changing geometry of decorations.} It appears that some notions from classical crystallography (like extinction rules and emergence of superstructure) may be also observed in quasicrystalline patterns. We concentrate here on the structure described in the Section \ref{sec:TheModel} with two decorating atoms. One of them is ascribed to sites $A$ and the second one to sites $B$, with the relative positions in respect to sites $a_{A}$ and $a_{B}$ respectively.

As it has been shown in reference \cite{Buczek2005a}, under special circumstances the quasicrystalline patterns can be periodic (both in the positions of the peaks and in their intensity), with the period being of integer multiplicity $q_{s}$, given by equation \eqref{eq:qs}. The necessary condition for this to happen is that for some values of $\fbr{n_{1},n_{2}}\in\Zahl^{2}$ vector $K$, given by equation \eqref{eq:K}, vanishes. Or, equivalently, there is a Bragg peak on position $k=n_{2}q_{s}$. Simple calculations show that such a situation is possible only in the case when $\kappa\in\Quotient$. The peaks occupying positions $k=n_{2}q_{s}$ are characterized by indexes $\fbr{n_{1},n_{2}}$ such that $n_{1}/n_{2} = 1/(\kappa - 1)$. These peaks are especially easy to investigate and we will focus our attention on them. 

\begin{figure}[htbp]
	\centering
		a)~\includegraphics[width=0.3\textwidth,origin=c,angle=-90]{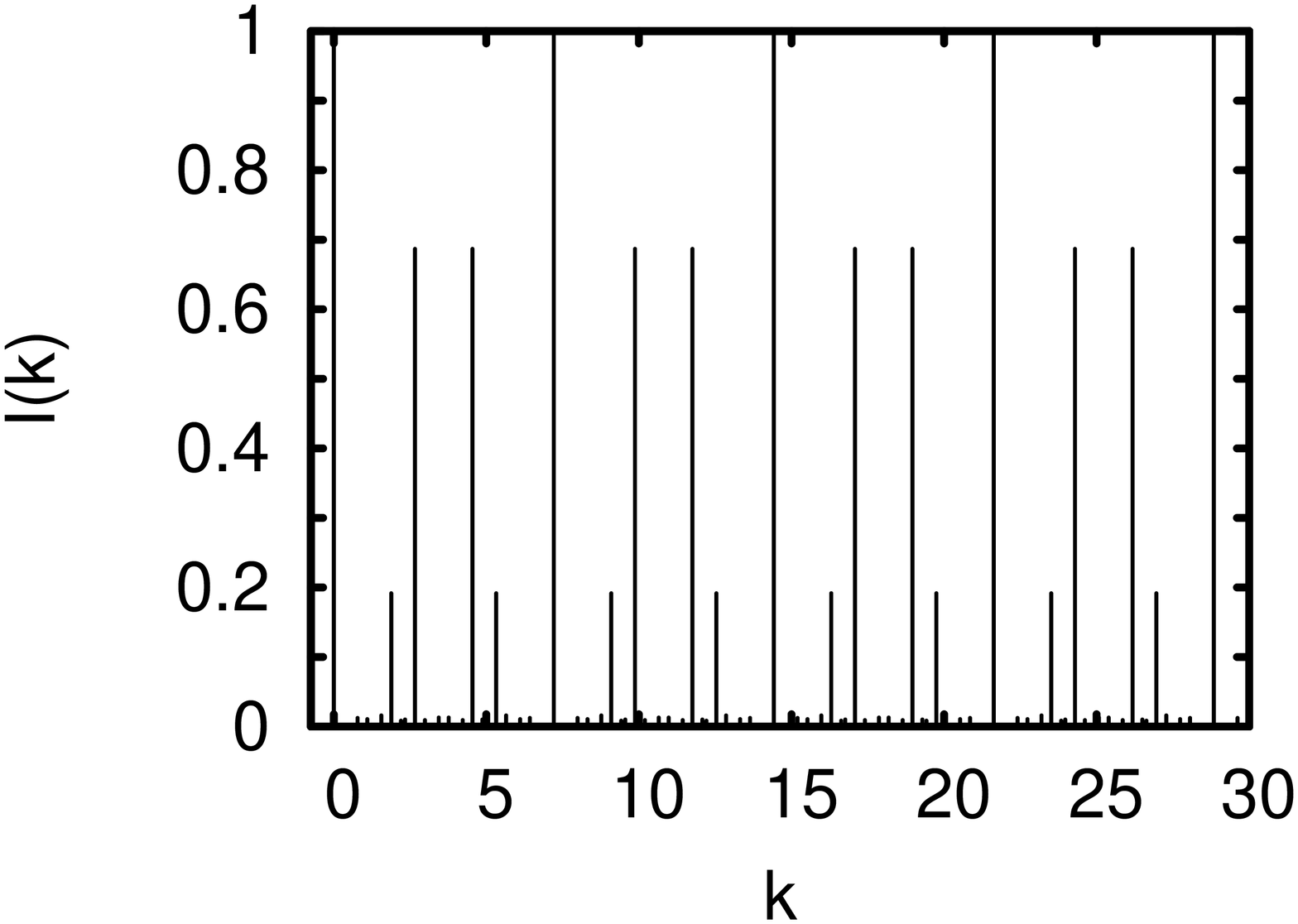}
		b)~\includegraphics[width=0.3\textwidth,origin=c,angle=-90]{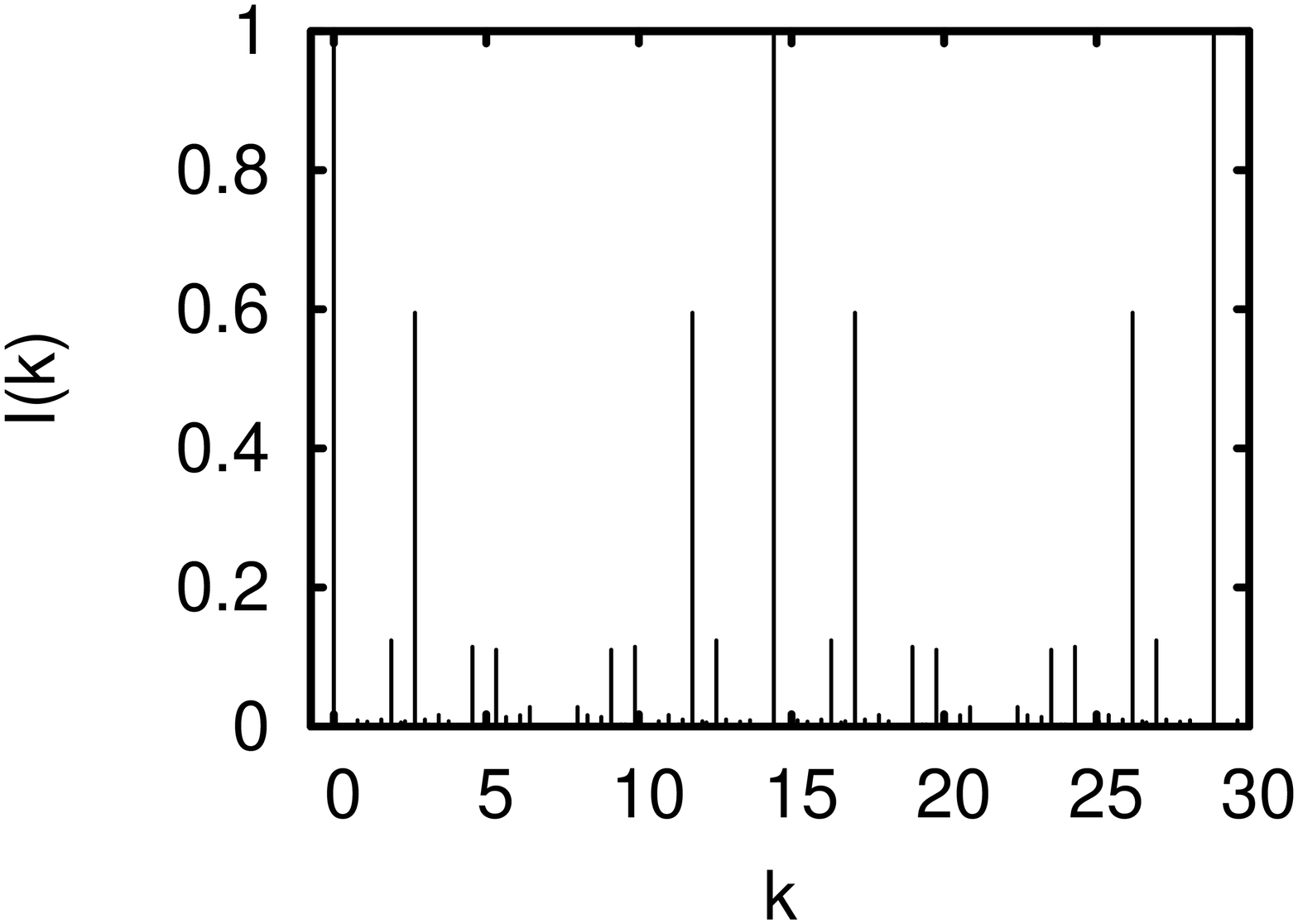}
	\caption{Examples of patterns with integer $c$ (please refer to the discussion in the text). Panel a) corresponds to $c = 2$, while panel b) for $c = 1$. The doubling of pattern period can be easily visible. In the case presented in b) every peak for $k = 2 m q_{s}$, $m\in\Zahl$ has intensity zero, please note however that intensity of other peaks also changes.}
	\label{fig:1}
\end{figure}

\begin{figure}[htbp]
	\centering
		a)~\includegraphics[width=0.3\textwidth,origin=c,angle=-90]{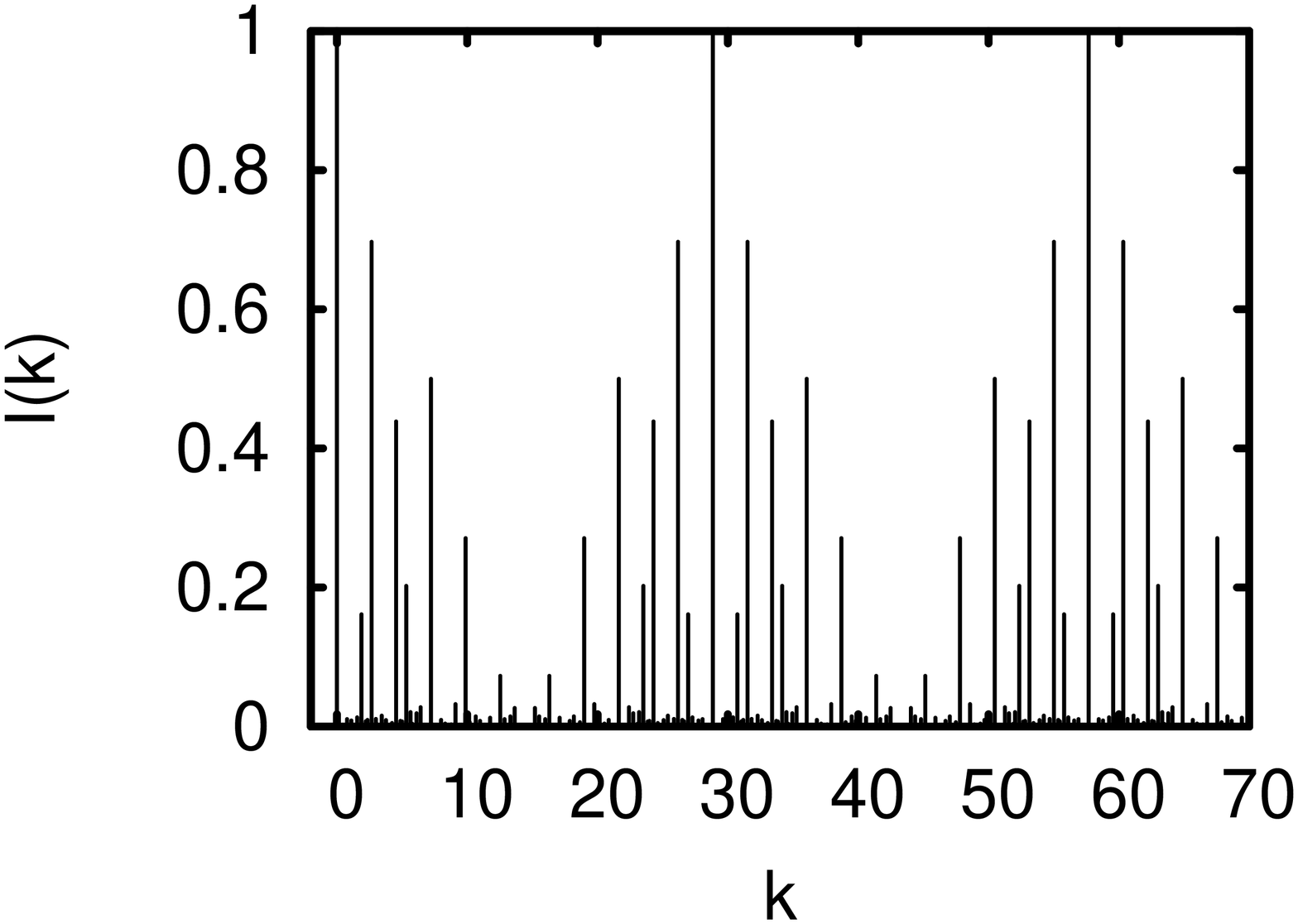}
		b)~\includegraphics[width=0.3\textwidth,origin=c,angle=-90]{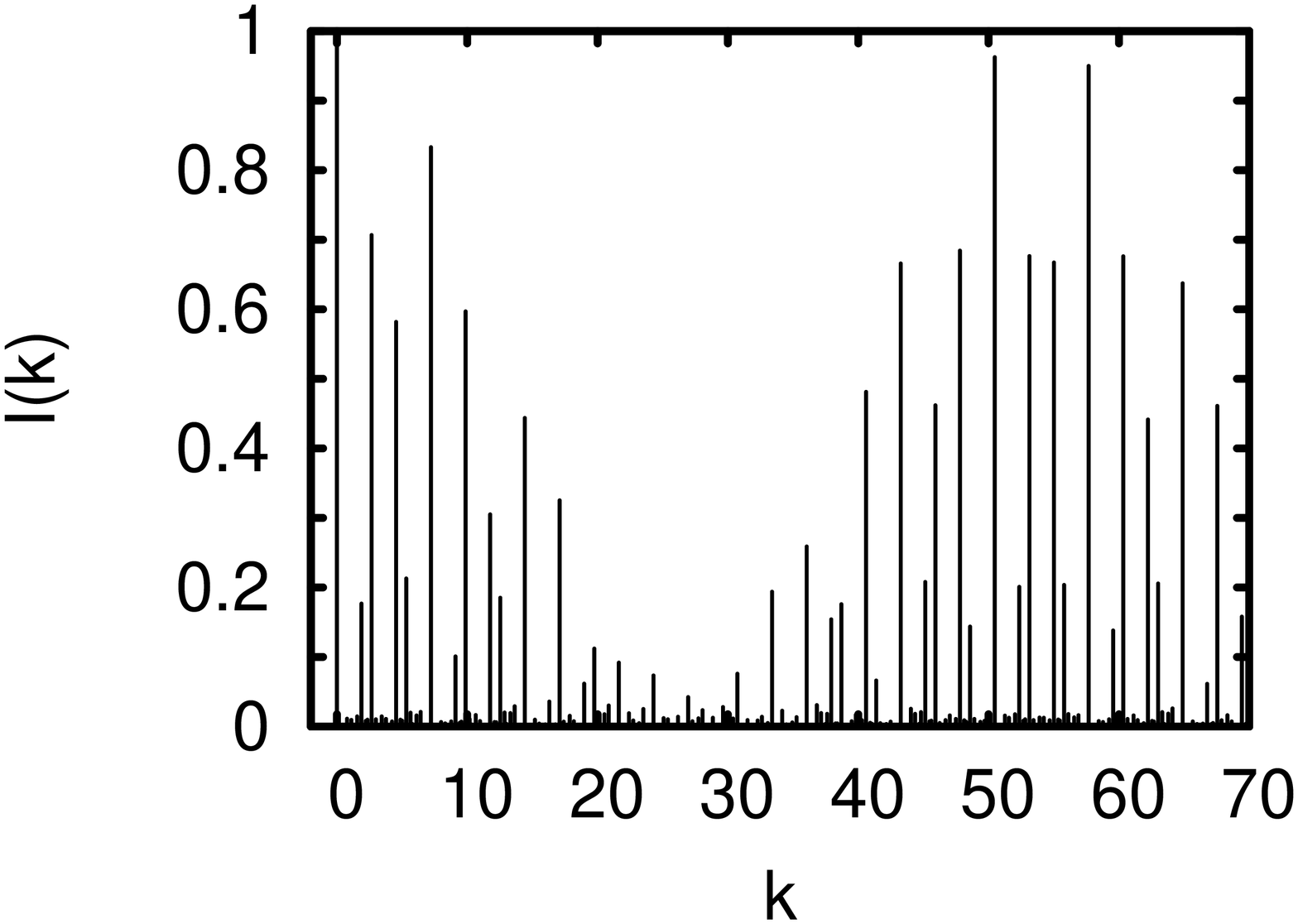}
	\caption{Examples of patterns with non-integer $c$.	Panel a) present the case for rational $c = 3/2$. Periodicity of the pattern is preserved, however the period may be high multiplicity of $q_{s}$ (here $4q_{s}$). For irrational $c$ (cf.~panel (b)) the pattern is again non-periodic (here $c=\sqrt{3})$.}
	\label{fig:2}
\end{figure}

We assume that $\kappa\in\Quotient$. The structure factor for peaks such that $k=n_{2}q_{s}$, according to \eqref{eq:GeneralStructureFactor}, is proportional to
\begin{align}
	\varphi_{A} f_{A} \exp(\ii n_{2} q_{s} a_{A}) + \varphi_{B} f_{B} \exp(\ii n_{2} q_{s} a_{B}).
\end{align}
Without any loss of generality we can take either $a_{A}$ or $a_{B}$ equal to arbitrary value and we take $a_{A} = 0$. We will vary position $a_{B}$. To make the total extinction of peaks $n_{2} q_{s}$ possible, we will set the scattering powers in such a way so $f_{X}\varphi_{X} = 1$. The structure factor becomes proportional to
\begin{align}
	1 + \exp(\ii n_{2} q_{s} a_{B}).
\end{align}
The analysis will now focus on the impact of parameter $a_{B}$ (the relative position of decorating atoms). Let us write
\begin{align}
	n_{2} q_{s} a_{B} = p \pi.
\end{align}
Parameter $p$ controls the extinctions. If $p$ is an even integer, peak on position $n_{2} q_{s}$ has intensity 1. When $p$ is odd, the peak vanishes. Since $a_{B}$ cannot depend on $n_{2}$ we must write $p = c n_{2}$ and therefore
\begin{align}
	a_{B} = c \frac{\pi}{q_{s}}.
\end{align}
$c$ is our control parameter. For even integer values of $c$ the diffraction pattern is periodic with period $q_{s}$, while for $c$ being odd the period is $2q_{s}$. In general for $c\in\Quotient$ the pattern is periodic with some integer multiplicity of $q_{s}$. The periodicity is once again destroyed for $c\in(\Reals\setminus\Quotient)$. Figures \ref{fig:1} and \ref{fig:2} present examples of diffraction patterns for structures with the following parameters $\sigma = 1/\sqrt{2}$ and $\kappa = 2$. This choice yields $k_{0} \approx 4.54656$, $q_{0} \approx 2.66331$ and $q_{s} \approx 7.20987$. 

\section{\clr{Summary}}
\label{sec:Summary}

\clr{Let us briefly summarize our observations. We have studied extinction rule for the 1D decorated quasicrystal induced by the change in the relative positions of decorating atoms. The exact and regular extinction rules are possible only for rational values of $\kappa$, i.e. in the case where the patterns are periodic (note however that the periodicity does not mean that the pattern is supported on the lattice). This observation applies however only to the case for which the atoms are placed exactly in the decoration sites. Once we allow them to be shifted with respect to their decoration sites the periodicity of the diffraction patterns might change or even be completely destroyed. Parameter $c$ that controls the behavior of patterns is the ratio of the relative displacement of the atoms from the sites, $a_{B} - a_{A}$, and the length scale introduced by vector $q_{s}$, $c= (a_{B} - a_{A}) q_{s}/\pi$. For rational $c$ the pattern is periodic (with period $m q_{s}$, $m\in\Zahl$). The periodicity is lost for irrational value of $c$ even for $\kappa\in\Quotient$. Let us also note that the presented extinction rule is general in the sense that it applies also for the structures with irrational $\kappa$. In this case however there are no peaks on the positions $k=n_{2}q_{s}$ that could be switched off by the presented extinction scheme.}

\bibliographystyle{acm} 
\bibliography{./quasi_bibliography}

\end{document}

%% file: headers.tex
\usepackage[T1]{fontenc}
\usepackage[latin2]{inputenc}
\usepackage[dvips]{graphicx}
\usepackage[mathcal]{eucal}
\usepackage{amsfonts}
\usepackage{amsmath}
\usepackage{amsthm}
\usepackage{amssymb}
\usepackage{bbm}
\usepackage{dsfont}
\usepackage{wasysym}  

%% file: commands.tex
\renewcommand{\vec}[1]
{
{\mathbf #1}
}

\newcommand{\indic}[1] 
{
{\mathds{1}_{#1}} 
}
\newcommand{\Heav} 
{
{\mathds{H}} 
}
\newcommand{\ii}
{
{{\mathbbm{i}}}   
}

\newcommand{\fbr}[1]
{
{\left( #1 \right)} 
}

\newcommand{\Reals}
{
{\mathbb{R}} 
}

\newcommand{\Zahl}
{
{\mathbb{Z}} 
}
\newcommand{\Quotient}
{
{\mathbb{Q}} 
}

\DeclareMathOperator{\floor}{floor}